\documentclass{vgtc}                          

\ifpdf
  \pdfoutput=1\relax                   
  \usepackage{graphicx}                
  \DeclareGraphicsExtensions{.pdf,.png,.jpg,.jpeg} 
\else
  \usepackage{graphicx}                
  \DeclareGraphicsExtensions{.eps}     
\fi%

\graphicspath{{figures/}{pictures/}{images/}{./}} 

\usepackage{microtype}                 
\PassOptionsToPackage{warn}{textcomp}  
\usepackage{textcomp}                  
\usepackage{mathptmx}                  
\usepackage{times}                     
\renewcommand*\ttdefault{txtt}         
\usepackage{cite}                      
\usepackage{tabu}                      
\usepackage{booktabs}                  
\usepackage{hyperref}

\onlineid{0}

\vgtccategory{Research}

\vgtcinsertpkg

\title{VisAnywhere: Developing Multi-platform\texorpdfstring{\\}{ }Scientific Visualization Applications}

\author{Thomas Marrinan\thanks{e-mail: tmarrinan@stthomas.edu; Also affiliated with ANL}, Madeleine Moeller\thanks{e-mail: maddie.moeller@stthomas.edu}, Alina Kanayinkal\thanks{e-mail: alina.kanayinkal@stthomas.edu}\\
    \parbox{3.2in}{\scriptsize \centering University of St. Thomas}
\and Victor A. Mateevitsi\thanks{e-mail: vmateevitsi@anl.gov}, Michael E. Papka\thanks{e-mail: papka@anl.gov; Also affiliated with UIC}\\
    \parbox{2.8in}{\scriptsize \centering Argonne National Laboratory}
}

\abstract{
Scientists often explore and analyze large-scale scientific simulation data by leveraging two- and three-dimensional visualizations. The data and tasks can be complex and therefore best supported using myriad display technologies, from mobile devices to large high-resolution display walls to virtual reality headsets. Using a simulation of neuron connections in the human brain, we present our work leveraging various web technologies to create a multi-platform scientific visualization application. Users can spread visualization and interaction across multiple devices to support flexible user interfaces and both co-located and remote collaboration. Drawing inspiration from responsive web design principles, this work demonstrates that a single codebase can be adapted to develop scientific visualization applications that operate everywhere.
} 

\CCScatlist{
  \CCScatTwelve{Human-centered computing}{Visu\-al\-iza\-tion}{Visu\-al\-iza\-tion application domains}{Scientific visualization};
  \CCScatTwelve{Human-centered computing}{Human computer interaction (HCI)}{Interaction paradigms}{Collaborative interaction};
  \CCScatTwelve{Computing methodologies}{Computer graphics}{Graphics systems and interfaces}{Virtual reality}
}

\begin{document}


\firstsection{Introduction}
\maketitle

Various display devices are best suited for different visualization tasks. Mobile devices like smartphones and tablets have led to higher user engagement and provide natural touch-based interaction \cite{Adepu_2016}. Large high-resolution display walls have led to improved sensemaking and naturally support co-located collaboration \cite{Czerwinski_2003}. Virtual reality environments have led to enhanced engagement and understanding of 3D objects \cite{Schuchardt_2007}. To support these various platforms, developers typically need to create separate applications specifically designed for each hardware/operating system.

Responsive visualizations \cite{Badam_2021}, however, enable visualizations to adapt to various displays. Using a simulation of brain plasticity \cite{Rinke_2016} with neuron positions based on real imaging from the Human Connectome Project \cite{Marcus_2011}, we showcase how web tools can be leveraged to develop multi-platform scientific visualization applications. Our solution also harnesses network communication, enabling co-located and remote collaboration using multiple devices simultaneously.

\section{Approach}
Our approach to develop a responsive and collaborative visualization application for brain plasticity involved data preprocessing, three-dimensional scientific visualization, two-dimensional information visualization, and multi-device collaborative interaction.

\subsection{Data Preprocessing}
Our primary objective when processing the simulation data was to enable interactive visualizations that could be accessed from any device with a standard internet connection. Uncompressed simulation data for four scenarios (no initial connectivity, learning, injury, and per-neuron calcium targets) was 117 GB and stored on a per-neuron basis. Given that our visualization aimed to illustrate brain activity at each timestep, displaying any one timestep would require loading all neuron data files for a given simulation scenario.

We therefore reorganized files to store data per-timestep and investigated more efficient methods for storing and transferring data to web applications. We chose to store data in the Parquet format \cite{Parquet} due to its lossless data compression and its provided JavaScript reader. We also aggregated connectivity on an area-to-area basis. This reduced data size and helped support creating an uncluttered visualization. After processing the raw simulation data, file sizes were reduced to a total of 38 GB (average of $<$ 1 MB per timestep).


\subsection{3D Visualization}
\begin{figure*}[tb]
 \centering 
 \includegraphics[width=\textwidth]{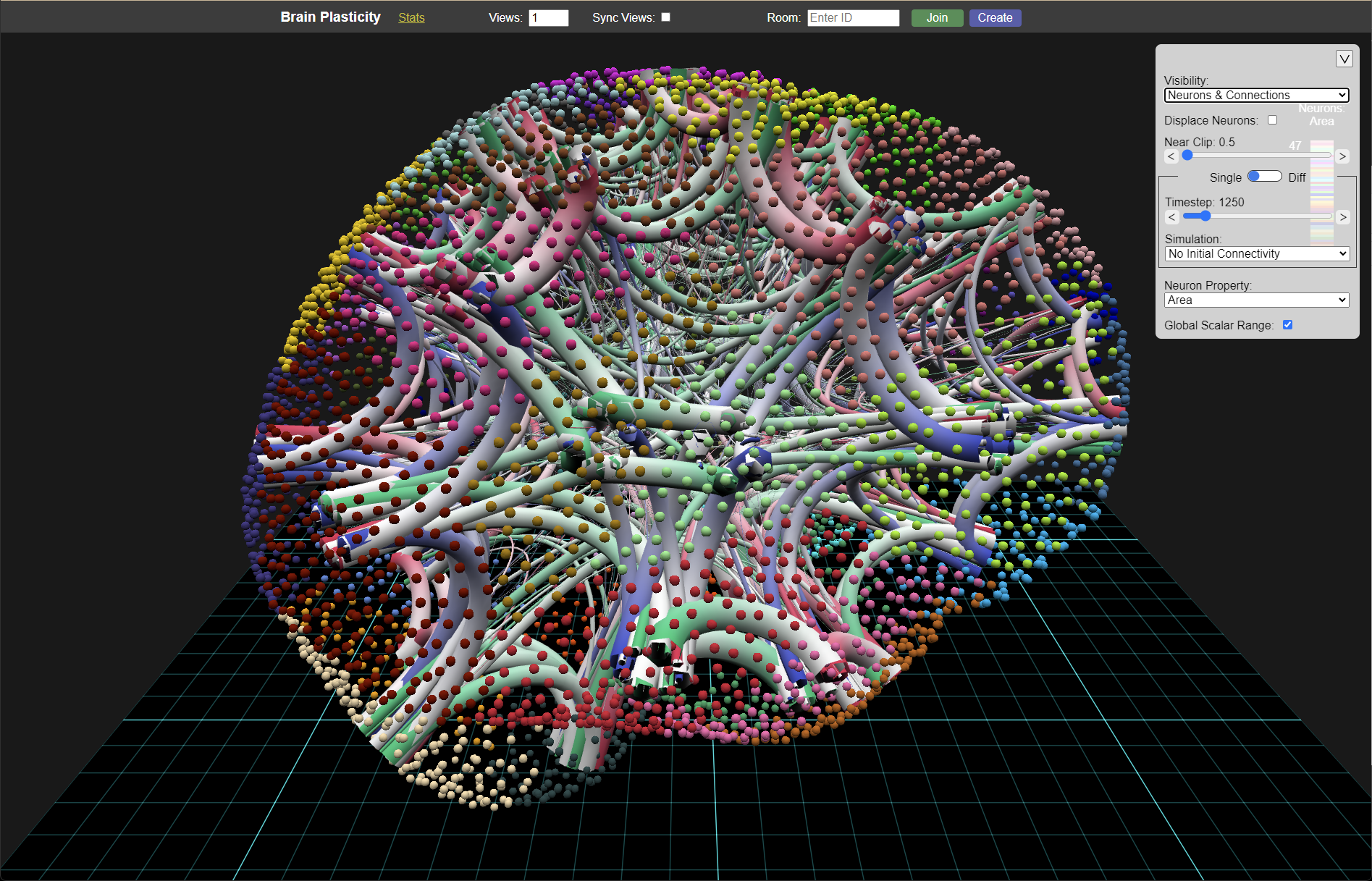}
 \caption{Overview of the brain plasticity 3D visualization application. The application shows neurons (spheres) and connections (tubes), and it contains a user interface for controlling the application.}
 \label{fig:vis3d_overview}
\end{figure*}

To create a 3D visualization of brain data that would run on mobile devices, desktop computers, large high-resolution display walls, and virtual reality headsets, we used Babylon.js \cite{BabylonJS} -- a web rendering engine. The visualization is comprised of two main components, the neurons and connections between neurons, plus a user interface (\autoref{fig:vis3d_overview}). Neurons are displayed as a sphere point-cloud, rendered using impostor spheres \cite{McKesson_2012}. The connections are displayed as tubes that follow a curved path joining two regions of the brain.

\subsubsection{Brain Visualization}
\begin{figure*}[tb]
 \centering 
 \includegraphics[width=\textwidth]{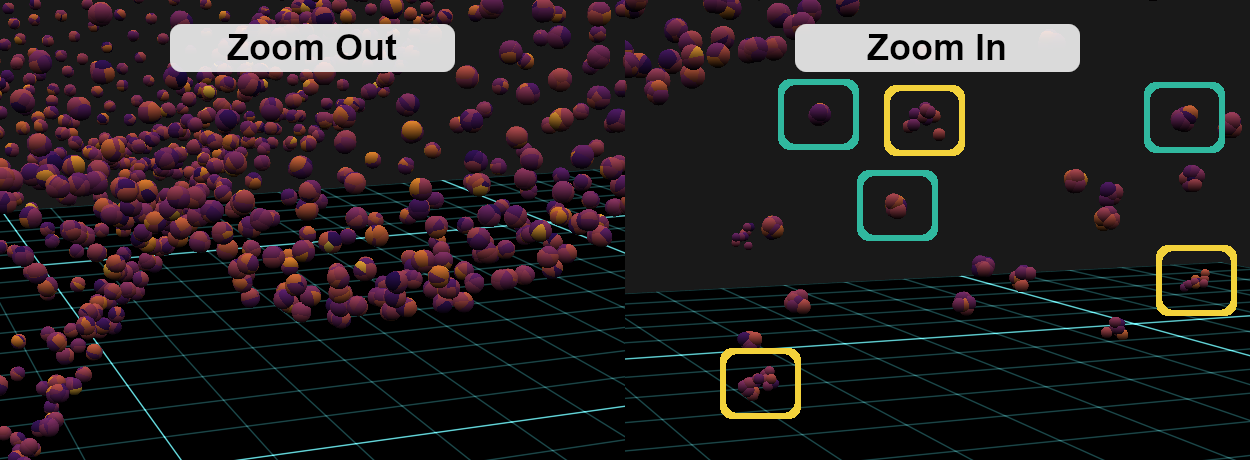}
 \caption{Dynamic radius of neuron spheres. Radius is set based on distance from the camera. This enables users to see individual neurons in nearby clusters while still being able to view far away neuron clusters. Turquoise highlighted areas show more distant neuron clusters where spheres are larger to remain visible but overlap each other. Yellow highlighted areas show nearby neuron clusters where each neuron is easily distinguishable.}
 \label{fig:dynamic_radius}
\end{figure*}

\begin{figure*}[tb]
 \centering 
 \includegraphics[width=\textwidth]{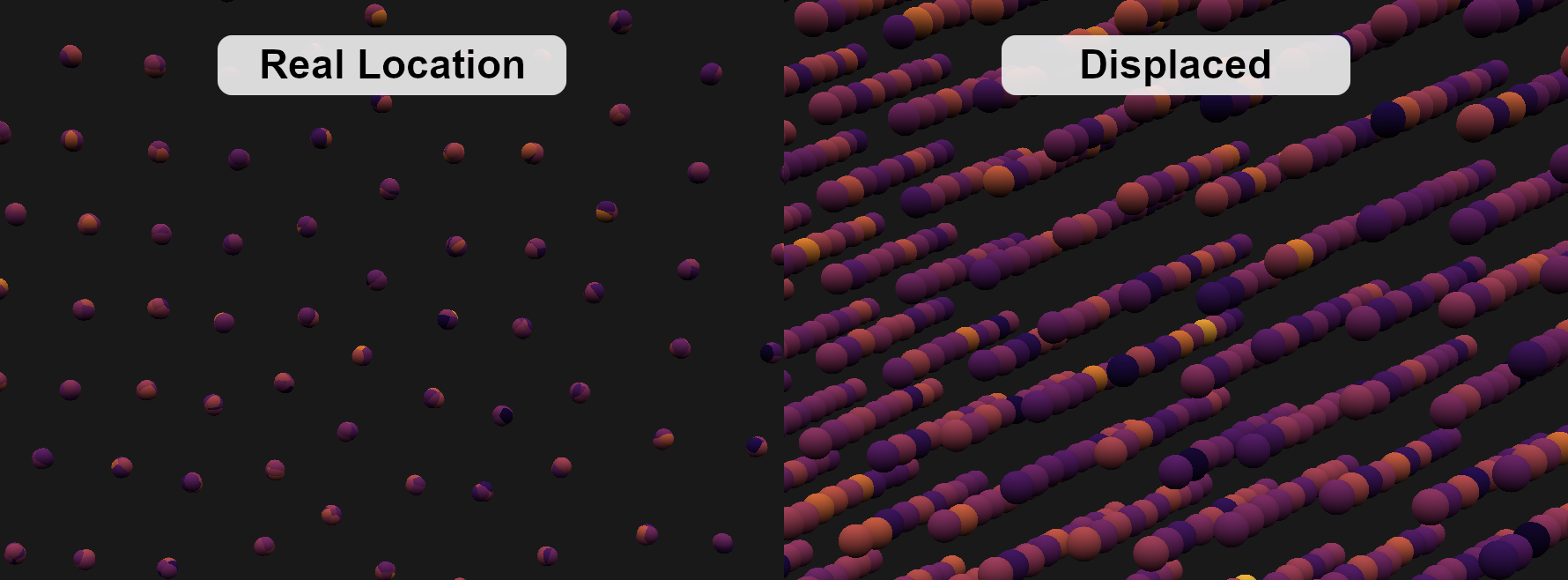}
 \caption{Neuron displacement. Neuron spheres can be displaced towards the center of the brain based on their index in each cluster of ten neurons. This prevents sphere overlap and creates a line of easily distinguishable spheres.}
 \label{fig:neuron_displace}
\end{figure*}

There are 50,000 neurons in the dataset, with 10 neurons generated at the 5,000 locations that are based on the underlying brain data (with each neuron have a slight random offset). Since clusters of 10 neurons are extremely close together, the spheres have significant overlap. In order to view the individual neurons, we provide two viewing options: A) dynamic sphere radius based on distance to camera, and B) displacing spheres towards the center of the brain. When viewing neurons using option A, the radius of nearby neurons decreases, thus separating the 10 neurons into easily distinguishable spheres while keeping further away neurons large enough to remain visible (\autoref{fig:dynamic_radius}). When viewing neurons using option B, the neurons area translated toward the center of the brain based on their index in the cluster, creating a line of easily distinguishable spheres (\autoref{fig:neuron_displace}).

\begin{figure*}[tb]
 \centering 
 \includegraphics[width=\textwidth]{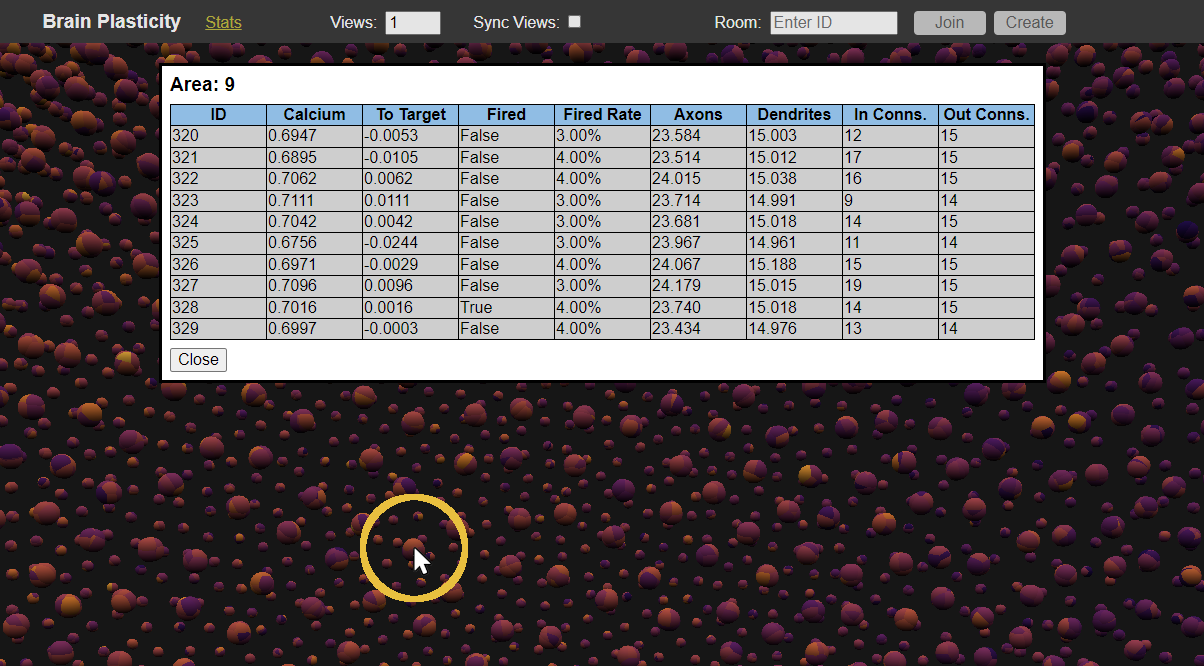}
 \caption{Table showing individual neuron properties. Users can select a cluster of neurons (ctrl + click) to pop up a table that contains raw values of all properties for the ten neurons in a cluster. Yellow highlight and mouse cursor show the neuron cluster that was selected.}
 \label{fig:neuron_selection}
\end{figure*}

Regardless of how the neurons are displayed, they can be colored based on simulation properties. Each neuron contains information on its area of the brain, amount of calcium, how far away the calcium level is from its target, whether the neuron fired or not on the current timestep, how frequently the neuron fired in the last 100 timesteps, number of axons, number of dendrites, number of outgoing synapses, and number of incoming synapses. The color can be adjusted using a global range (min/max for the entire simulation) or a local range (min/max for the given timestep). Users can also investigate individual neurons -- selecting a cluster of 10 neurons will pop up a table with raw values for each neuron property (\autoref{fig:neuron_selection}). 

\begin{figure*}[tb]
 \centering 
 \includegraphics[width=\textwidth]{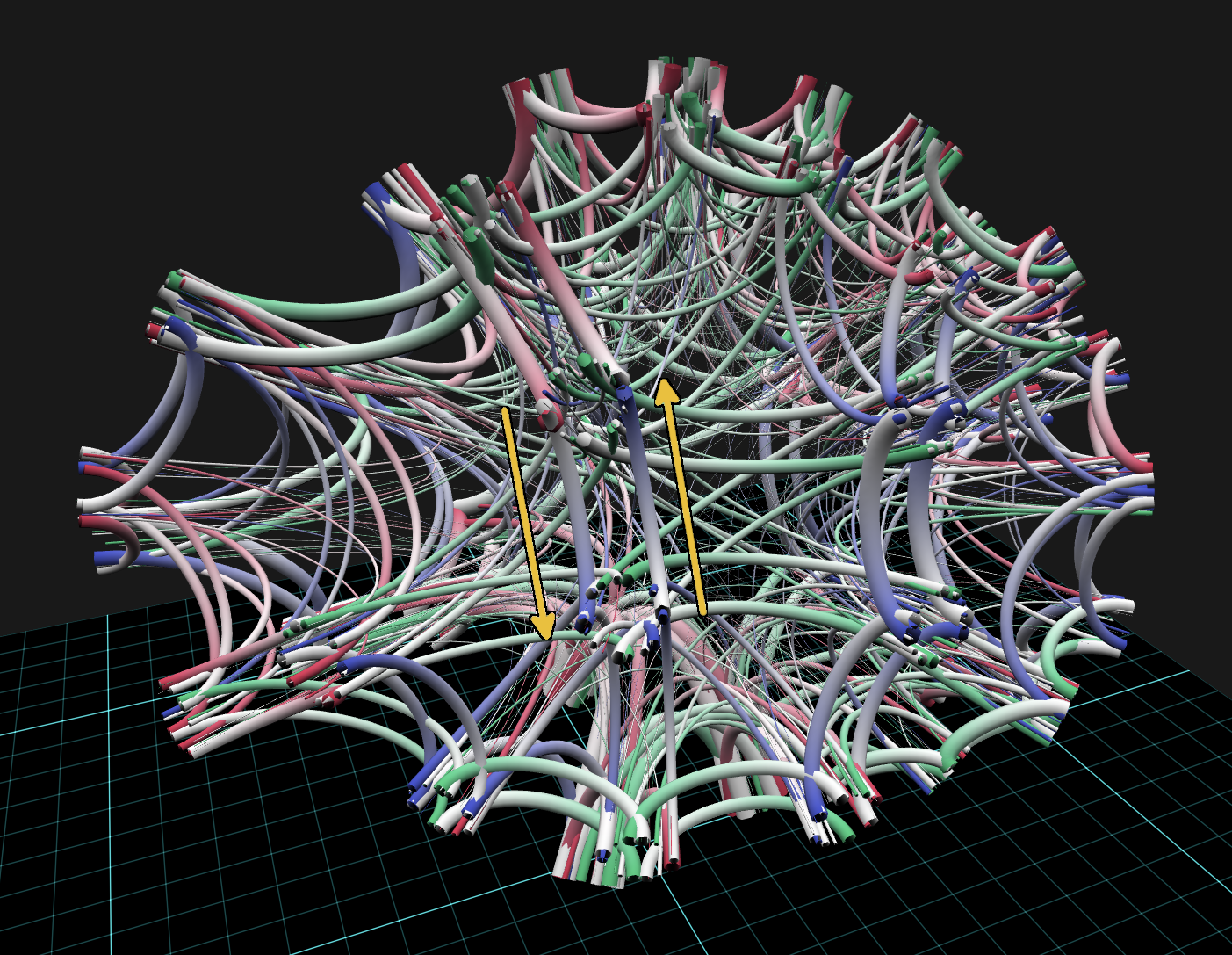}
 \caption{Tubes showing area-to-area connections. All tubes curve towards the center of the brain so as to ensure all connections remain on the interior. The radius of each tube corresponds to the number of connections from one area to another. The color indicates primary connection direction (red: left/right, green: back/front, blue: bottom/top), similar to DTI connectome imaging. The color gradient depicts connection direction (from white to colored), as highlighted by the yellow arrows.}
 \label{fig:brain_connections}
\end{figure*}

Connections are visualized as tubes joining one area of the brain to another (\autoref{fig:brain_connections}). The radius of each tube represents the number of connections between neurons in the two adjoining areas. The tubes are colored based on primary connection direction: (red: left/right, green: back/front, blue: bottom/top). Each tube's color has a gradient, from white to colored, that indicated connection direction.

\begin{figure*}[tb]
 \centering 
 \includegraphics[width=\textwidth]{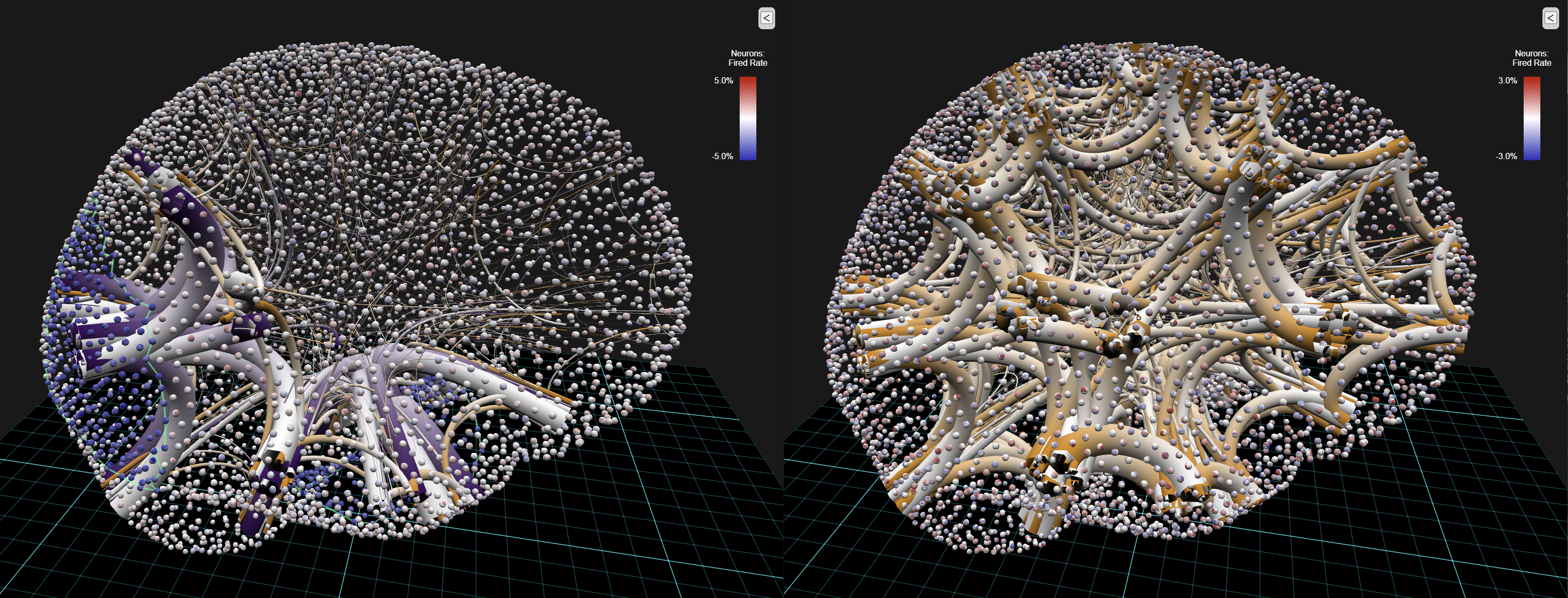}
 \caption{Difference visualization. Neuron colors show change from one dataset to another (blue: property value decreased, white: property value stayed the same, red: property value increased). Connection radius represents the absolute value of the change in number of connections between areas, and color represents whether connections were gained (orange) or lost (purple). Left panel shows the difference in the brain injury simulation, from 100 output timesteps prior to injury to 100 output timesteps after injury. Right panel shows the difference between the no initial connectivity and learning simulations, both on the same timestep (10\% of the way through the simulation).}
 \label{fig:diff}
\end{figure*}

Both neuron and connection data can additionally show differences between two pieces of data. This enables users to easily visualize the change between timesteps or simulation scenario (\autoref{fig:diff}). When visualizing differences, neuron colors always use a local scale and the connection tube radius is based upon the absolute value of the change, with tube color representing whether the change was positive (orange) or negative (purple).

\begin{figure*}[tb]
 \centering 
 \includegraphics[width=\textwidth]{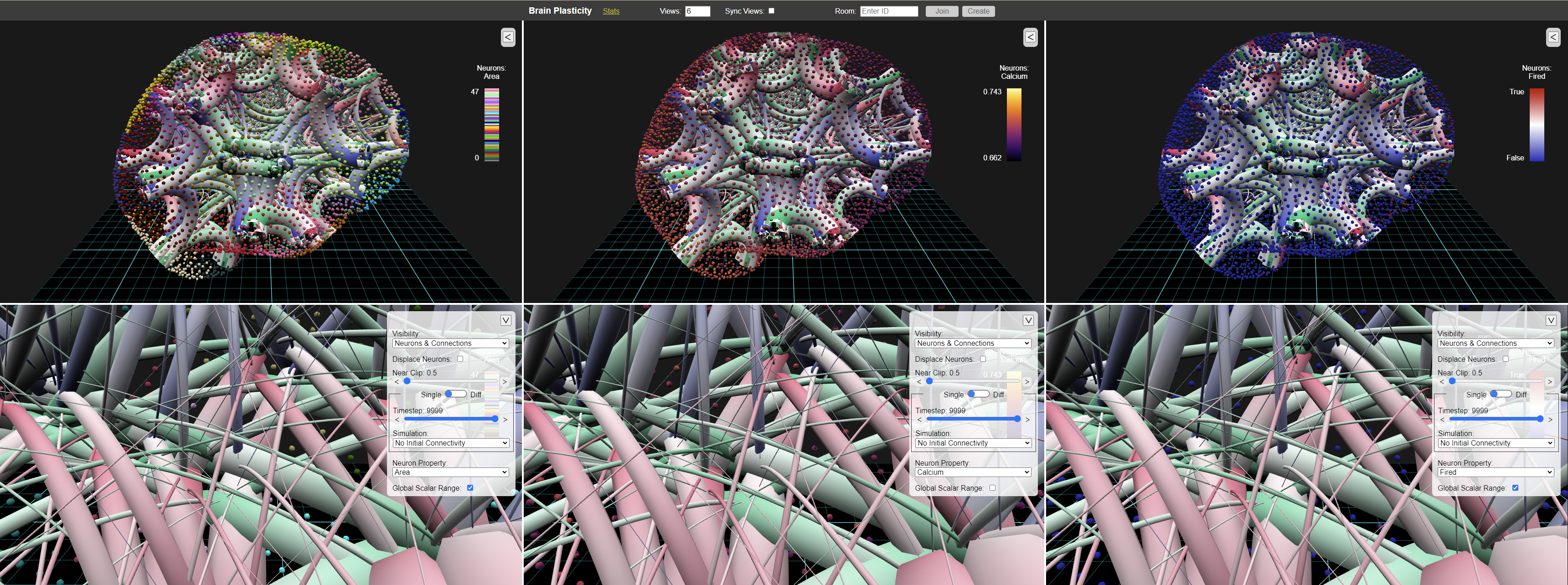}
 \caption{Multiple views to support ensemble comparison. Data visualized in each view is controlled independently. The camera can also be individually controlled for each view, or they can be synchronized across all views.}
 \label{fig:multiple_views}
\end{figure*}

Our application supports showing 1-8 views of the brain visualization. The number of supported views is based on the display device (e.g. smartphones are limited to one view, whereas large high-resolution display walls can show up to eight views). Each view is independently controlled, enabling users to compare multiple timesteps, simulation scenarios, or neuron properties (\autoref{fig:multiple_views}).

\subsubsection{User Interface}
The user interface is broken into two parts, a global user interface that controls entire application settings and local user interfaces that control settings for each individual 3D view. The global user interface lets users choose how many 3D views to show, whether the views should be synchronized, and controls for creating/joining a collaboration session. The maximum number of views supported is based on display resolution, thus ensuring sufficient screen real-estate for each view. Enabling view synchronization will cause the camera position and orientation to remain uniform for all views, making it easy for users to look at the same region of the brain in multiple views. Collaboration sessions enable data to synchronize across multiple devices, thereby supporting remote collaboration and multi-device visualization (e.g. 3D and 2D visualization apps).

Local user interfaces enable users to control the 3D visualization in each view. Various widgets allow users to select data visibility (neurons, connections, or both), whether to visualize neurons with dynamic radius or displace them towards the center of the brain, near clipping plane distance, whether to visualize raw data or the difference between two datasets, simulation timestep, simulation scenario, property by which to color neurons, and whether to color neurons based on the global or local range.

Global and local user interfaces can be seen in \autoref{fig:vis3d_overview}.

\subsection{2D Visualization}
\begin{figure*}[tb]
 \centering 
 \includegraphics[width=\textwidth]{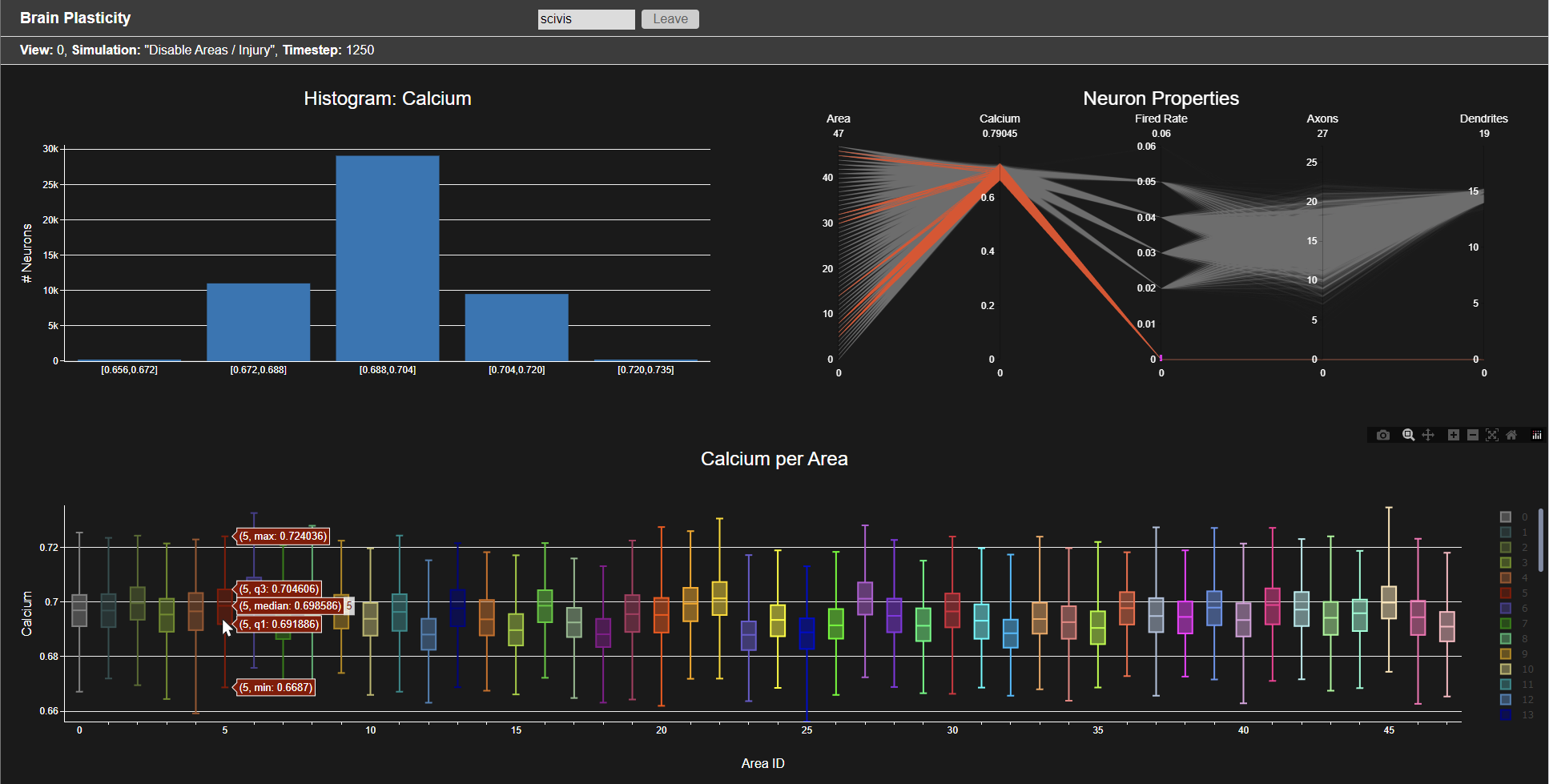}
 \caption{2D charts that visualize statistics about the brain plasticity data. Top-left chart shows a histogram of the selected neuron property. Top-right chart shows a parallel coordinates graph for area, calcium, fired rate, number of axons, and number of dendrites. Bottom chart shows a box and whisker plot for the selected neuron property, separated by brain area. The charts are all interactive -- for example, the parallel coordinates graph can be filtered on any axis and hovering over an area in the box and whisker plot will show numeric values. Charts are also responsive -- the three charts will be stacked on top of one another on devices with lower resolution.}
 \label{fig:charts2d}
\end{figure*}

\begin{figure*}[tb]
 \centering 
 \includegraphics[width=\textwidth]{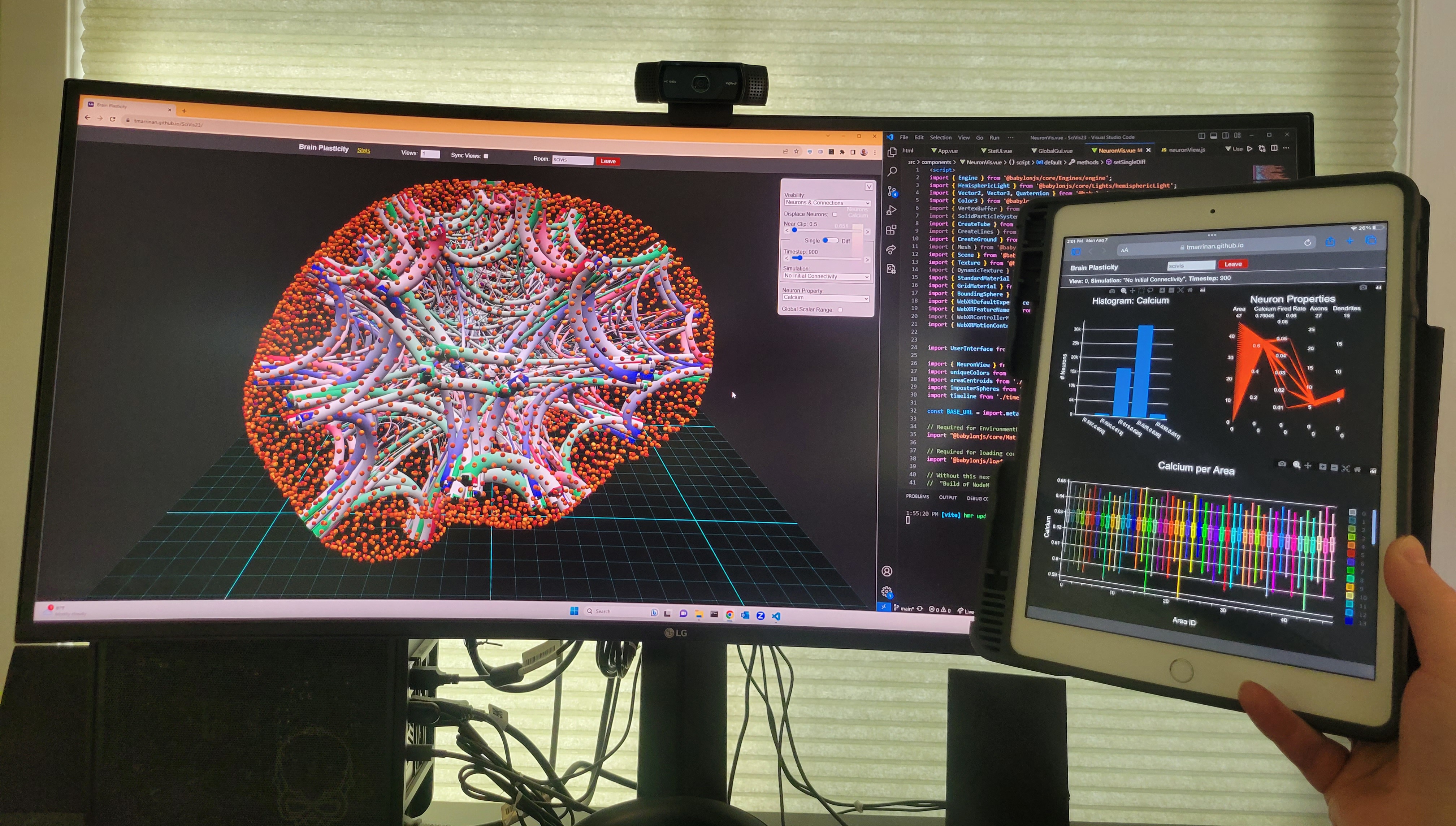}
 \caption{3D and 2D visualizations synchronized via collaboration session to support multi-device viewing. The charts shown in the 2D visualization automatically update based on the data viewed in the 3D visualization.}
 \label{fig:multidevice_3d2d}
\end{figure*}

In order to provide a statistical overview of the simulation data and to help illuminate underlying patterns, we created a second web application meant to be used in conjunction with the 3D visualization application. This application consists of three 2D charts made using Plotly \cite{PlotlyJS} (\autoref{fig:charts2d}). The first chart shows a histogram of the currently viewed neuron property. The second chart shows a parallel coordinates graph with 5 axes (area, calcium, fired rate, number of axons, and number of dendrites). The third chart shows a box and whisker plot of the currently viewed neuron property, separated by brain area. The 2D charts are synchronized to the 3D visualization via collaboration session (\autoref{fig:multidevice_3d2d}). Since we used Plotly, the charts also support user interaction to further analyze or filter the data.

\subsection{Virtual Reality}
\begin{figure*}[tb]
 \centering 
 \includegraphics[width=\textwidth]{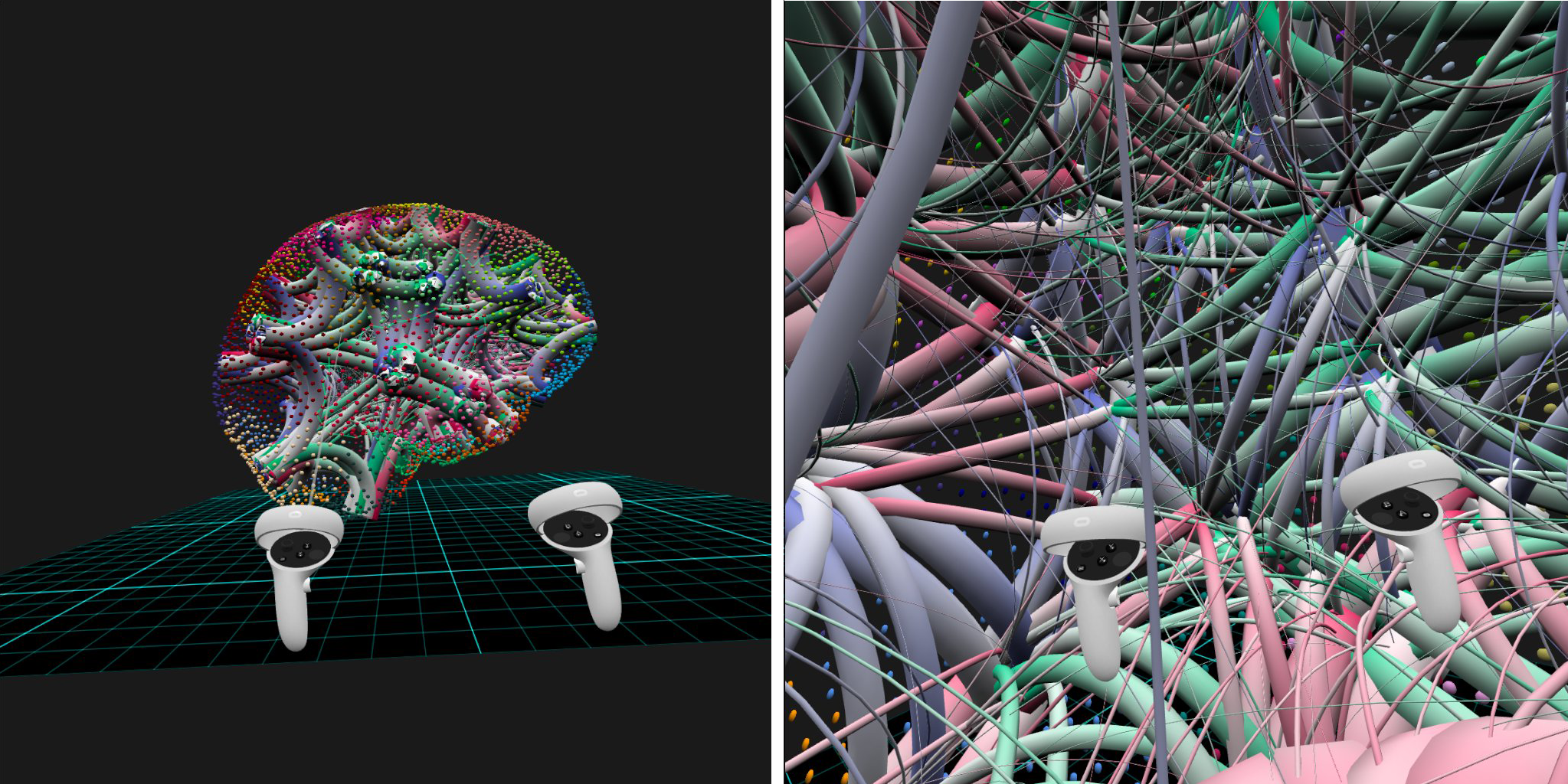}
 \caption{Virtual reality visualization as seen from inside a VR headset. Left image displays view from outside the brain, while right image shows a view from inside the brain.}
 \label{fig:brain_vr}
\end{figure*}

Exploring a three-dimensional dataset using a virtual reality (VR) headset provides an immersive experience with stereoscopic depth cues. We therefore leveraged WebXR to enable immersive viewing of our 3D visualization on any VR-capable device. Users can seamlessly jump between the standard 3D visualization and immersive VR mode, allowing users to select a dataset of interest in the standard view before transitioning into VR to explore the data further. The VR view can also be synchronized via collaboration session. Once in immersive VR mode, users can utilize VR controllers to navigate within the visual representation of the brain (\autoref{fig:brain_vr}).

\section{Results}
\begin{figure*}[tb]
 \centering 
 \includegraphics[width=\textwidth]{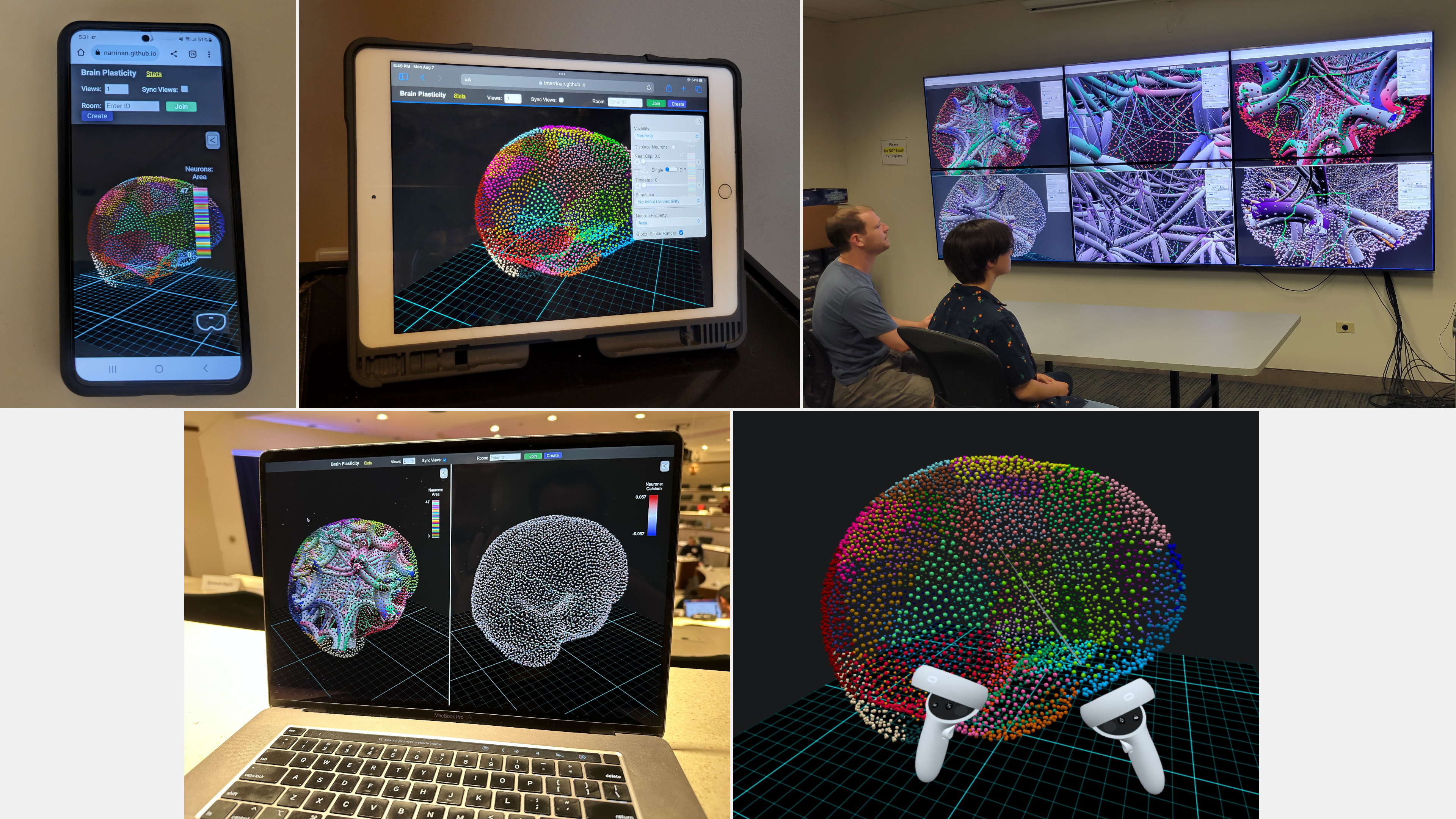}
 \caption{Multi-platform scientific visualization application. The brain plasticity application is shown running on a smartphone, tablet, laptop, large high-resolution display wall, and virtual reality headset.}
 \label{fig:vis3d_devices}
\end{figure*}

We successfully developed a multi-platform scientific visualization application. Web technologies enabled a single codebase to run anywhere without needing special consideration for various device types or operating systems (\autoref{fig:vis3d_devices}). Our brain plasticity application is available at \url{https://tmarrinan.github.io/SciVis23/}.

\subsection{Evaluation with Domain Experts}
We evaluated a prototype of our visualization by conducting informal interviews with two neuroscientists. Both of them found the application highly intuitive and expressed considerable enthusiasm for the project. They could interpret the data immediately without requiring an explanation of what was being visualized. They requested the capability to compare differences between datasets and the functionality to graph variables such as calcium levels. This motivated our subsequent implementation of these features. Both scientists agreed to continued collaboration and expressed a strong desire to incorporate our tool into their workflow.

\subsection{Achieved Tasks}
The main 3D visualization enables users to explore various datasets. Supporting multiple views allows users to view the same dataset from multiple perspectives or compare different datasets. The ability to visualize differences between datasets, as well as the supplementary 2D charts, help users analyze the relation between various neuron properties and connection creation/deletion events. These features are combined with a unified interface that allows individuals or teams to intuitively investigate brain plasticity simulations.

\acknowledgments{
This research was supported in part by the Argonne Leadership Computing Facility, which is a U.S. Department of Energy Office of Science User Facility operated under contract DE-AC02-06CH11357. We would also like to thank domain experts Narayanan `Bobby' Kasthuri (ANL, UChicago) and Kevin Michael Boergens (UIC) for their time, help, and insights.}

\renewcommand*{\ttdefault}{PTSansNarrow-TLF}
\urlstyle{tt}

\bibliographystyle{abbrv-doi-hyperref-narrow}
\bibliography{references}

\renewcommand*{\ttdefault}{txtt}
\pagebreak

\end{document}